\documentstyle[11pt,newpasp,twoside,epsf]{article}
\begin{document}

\title{X-ray emitting AGN unveiled by the Chandra Multiwavelength Project}

\author{John D. Silverman\altaffilmark{1},Paul J. Green, Thomas
L. Aldcroft, Dong-Woo Kim, Wayne Barkhouse, Robert A. Cameron, Belinda
J. Wilkes} \affil{Harvard-Smithsonian Center for
Astrophysics, 60 Garden Street , Cambridge, MA 02138}


\altaffiltext{1}{jsilverman@cfa.harvard.edu}

\begin{abstract}

We present an X-ray and optical analysis of a flux limited
(f$_{2.0-8.0 \rm{keV}} > 10^{-14}$ erg s$^{-1}$ cm$^{-2}$) sample of
126 AGN detected in 16 Chandra fields.  This work represents a small
though significant subset of the Chandra Multiwavelength Project
(ChaMP).  We have chosen this limiting flux to have a reasonable
degree of completeness (50\%) in our optical spectroscopic
identifications.  The optical counterparts of these hard AGN
are characterized as either broad emission line AGN (BLAGN; 67\%),
narrow emission line galaxies (NELG; 22\%) or absorption line galaxies
(ALG; 11\%) without any evidence of an AGN signature.  Based on their
X-ray luminosity and spectral properties, we show that NELG and ALG
are primarily the hosts of obscured AGN with an intrinsic absorbing
column in the range of $10^{21.5}<$ N$_{\rm{H}}<10^{23.3}$ cm$^{-2}$.
While most of the BLAGN are unobscured, there are a few with
substantial absorption.  X-ray surveys such as the ChaMP nicely
complement optical surveys such as the SDSS to completely determine
the demographics of the AGN population.

\end{abstract}

\section{Introduction}

X-ray surveys of the extragalactic universe in the era of Chandra and
XMM-Newton are for the first time able to probe the demographics of
the AGN population irrespective of any moderate obscuration.  Current
deep surveys such as the CDF-N (Brandt et al. 2002) and the CDF-S
(Tozzi et al. 2001) are unveiling AGN with an abundant amount of gas
hiding the bright quasars and the lower luminosity Seyfert galaxies.
This obscuration can be large enough to effectively hide any optical
signature of an active nucleus and prevent the inclusion of these
sources in optical surveys such as the SDSS.  With the unprecedented
sensitivity and resolving power of these current observatories, we are
able to probe large volumes in an unbiased fashion to determine the
prevalence of X-ray emitting AGN and their subsequent evolution.

How do these obscured sources fit into the AGN unification scheme?
Many of these do not necessarily have optical AGN signatures.  Is this
a result of host dilution (Moran, Filippenko, \& Chornock 2002) or
some other geometry/structure that prevents us from viewing the narrow
line emitting gas?  While optical extinction and X-ray absorption
normally go hand in hand (Turner et al. 1997), there are a number of
cases to the contrary (i.e. Akylas, Georgantopoulas, \& Barcons 2003;
Panessa \& Bassani 2002).

While the deep fields do cover a large volume, shallower and wide
field surveys are needed to compile a significant sample of sources
with 2-8 keV flux levels around $10^{-14} - 10^{-15}$ erg s$^{-1}$
cm$^{-2}$ which comprise most of the 2-8 keV CXRB (Cowie et al. 2002).
With large samples of all AGN types, we can determine the relative
importance and nature of these new AGN to the parent population.

\section{The Chandra Multiwavelength Project (ChaMP)}

The ChaMP provides a medium-depth, wide-area sample of serendipitous
X-ray sources from archival Chandra fields covering $\sim 14$ deg$^2$.
The broadband sensitivity between 0.3--8.0 keV enables the selection
to be far less affected by absorption than previous optical, UV, or
soft X-ray surveys.  Chandra's small point spread function
($\sim$1$\arcsec$ resolution on-axis) and low background allow sources
to be detected to fainter flux levels.  The project effectively
bridges the gap between flux limits achieved with the Chandra deep
field observations and those of past ROSAT and ASCA surveys.

We present preliminary results from the ChaMP using a bright subsample
(126 identified AGNs; f$_{2.0-8.0\rm{keV}} > 10^{-14}$ erg s$^{-1}$
cm$^{-2}$) of 437 hard X-ray sources detected in 16 ChaMP fields.
This flux limit includes a significant fraction of sources with
spectroscopic identifications.  To construct a pure AGN sample, we
require the observed 2.0-8.0 keV luminosity to exceed 10$^{42}$ erg
s$^{-1}$.  Our motivation is to determine the demographics of the hard
X-ray emitting AGN, measure the range of intrinsic obscuration, and
determine the extent to which obscuration of X-rays translates to
extinction in the optical.

\subsection{X-ray observations}

We have chosen 16 Chandra fields for which we have acquired followup
optical imaging and spectroscopy.  A full description of the ChaMP
image reduction and analysis pipeline XPIPE can be found in Kim et
al. (2003).  For the following analysis, we require a S/N $>$ 2 in the
2.5-8.0 keV band to generate a hard X-ray AGN catalog which minimizes
any inherent systematics due to absorption of soft X-rays.  We
restrict the off-axis angle of the Chandra detections to less than
$12\arcmin$ since the sensitivity is significantly reduced.

\subsection{Optical followup}

We have acquired optical imaging for each Chandra field to identify
counterparts to X-ray sources.  We have utilized the NOAO 4m
telescopes with the MOSAIC camera to cover the full Chandra field of
view.  The complete details of the ChaMP optical followup program
including strategy, image reduction, and source detection can be found
in Green et al. (2004).

Optical spectroscopy is crucial for determining the source type and
redshift.  The majority of optical spectra are acquired from the
CTIO/4m and WIYN/3.5m with HYDRA, a multi-fiber spectrograph.  To
extend the identifications beyond r$^{\prime}\sim21$, we are observing
on Magellan and the MMT to reach r$^{\prime}\sim22.5$.  Objects with
strong emission lines (W$_{\lambda}>5 \rm{\AA}$) are classified as
either Broad Line AGN (BLAGN; FWHM $>$ 1000 km s$^{-1}$) or Narrow
Emission Line AGN (NELG).  For counterparts with weak emission line or
purely absorption line, spectra are classified as Absorption Line
Galaxies (ALG).  Any stellar source is labelled as STAR.  If the
associated X-ray emission is extended the object is further labelled
as a cluster member.

\section{ChaMP AGN: X-ray and optical properties}

We show the optical magnitude as a function of X-ray flux for all
sources (378) detected in 14 of the 16 fields which have reliable
optical photometry (Fig. 1).  We find that 67\% of the identified AGN
are classified as BLAGN.  Many of the Chandra sources (NELG-22\%,
ALG-11\%) do not resemble the typical AGN found in optical surveys.
Their spectrum is characteristic of the host galaxy and not primarily
associated with the AGN itself.

\begin{figure}
\plottwo{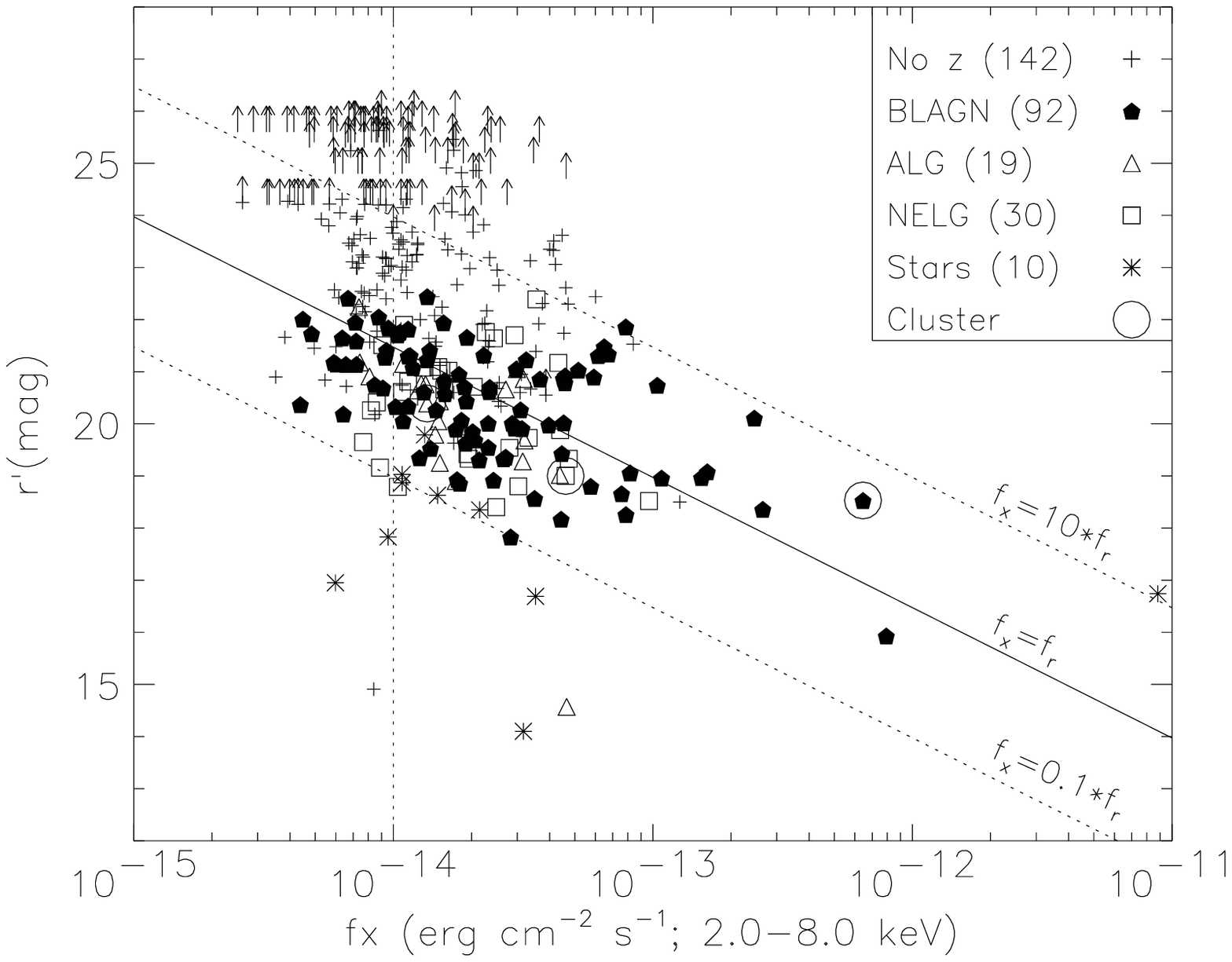}{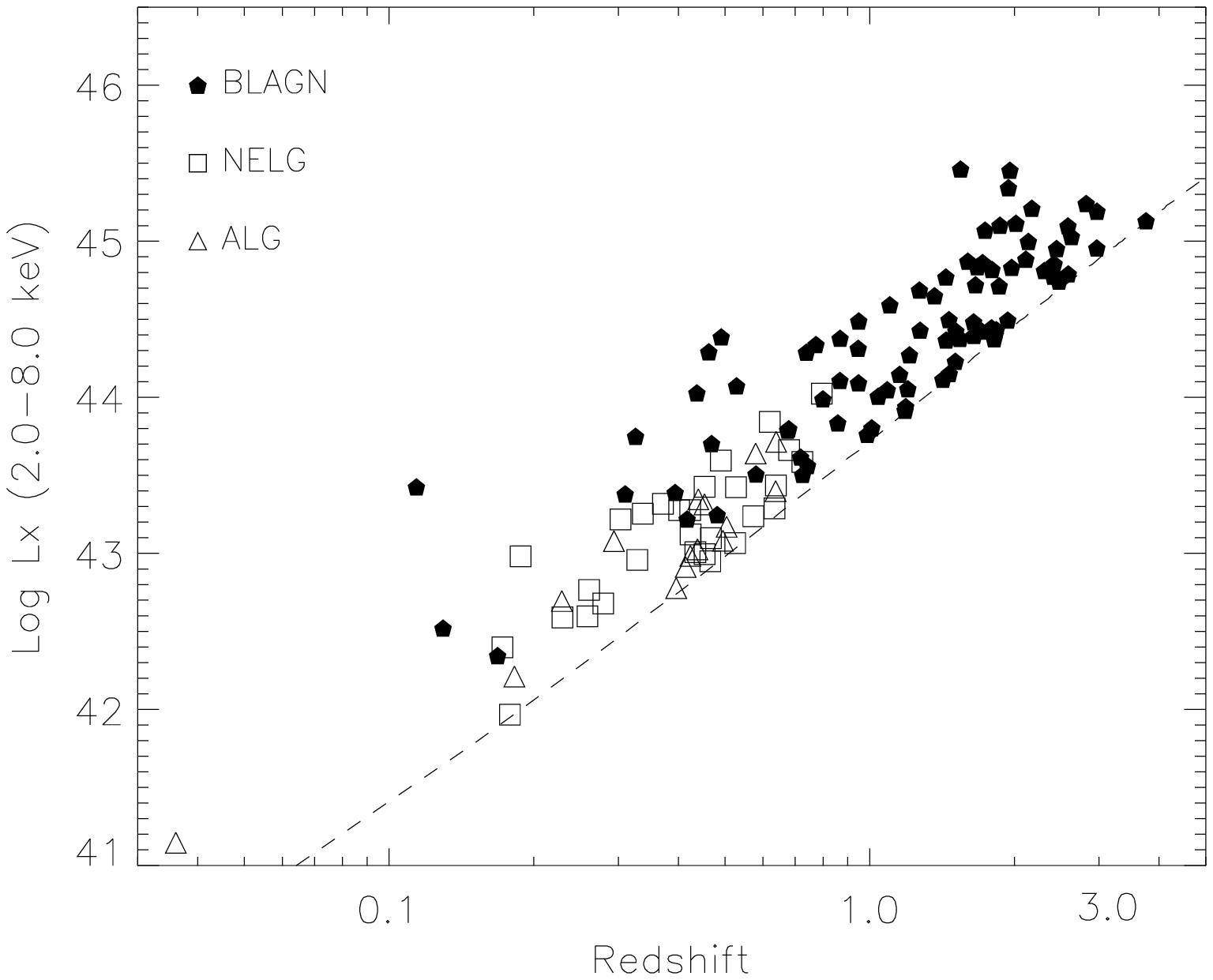}
\caption{{\it left} X-ray flux (2-8 keV) vs. optical magnitude (r$^{\prime}$).
{\it right} Luminosity (erg s$^{-1}$)/redshift distribution of ChaMP AGN.}
\end{figure}

Since 85\% of all Chandra sources in these medium depth fields are
AGN, hard X-ray surveys have a high degree of efficiency for finding
accreting, supermassive black holes.  It is evident that Chandra is
capable of detecting hard AGN out to z$\sim$4 (Fig. 1).  Based on the
strong X-ray luminosity of the NELG and ALG, we would expect to detect
optical emission from the AGN which suggests the presence of severe
extinction.

The NELG and ALG are only seen for z$<$0.8.  The steep drop in their
numbers for z$>0.8$ is primarily due to a selection bias.  A luminous
galaxy (10L$_{\star}$) at z$\sim$0.8 is fainter than our limit for
optical spectroscopic followup (r$^{\prime}$=22).

\section{X-ray spectral fitting and intrinsic absorption}

We find that there is a direct relationship between the X-ray and
optical properties of these hard AGN. We fit the X-ray count
distribution for each source with a powerlaw with the spectral index
($\Gamma$; Fig 2 left.) and N$_{\rm{H}}$ set to the galactic value.
Errors are 90\% confidence intervals.  The BLAGN have $\Gamma\sim1.9$
which is expected for unabsorbed AGN.  Most NELG and ALG have a
flatter spectral slope ($\Gamma<1.5$).  The AGN that lack broad
optical emission lines, probably due to dust extinction, suffer from
significant X-ray absorption as well.

We re-ran our fitting routine, assuming all sources can be fit with a
powerlaw, fixed $\Gamma=1.9$, galactic N$_{\rm{H}}$ and an additional
intrinsic absorber.  We find that the NELG and ALG have intrinsic
absorbing columns with $10^{21.5}<$ N$_{\rm{H}}<10^{23.3}$ cm$^{-2}$
(Fig. 2 right).  While most BLAGN are consistent with no intrinsic
absorption, there exist a few BLAGN which are absorbed.  As evident,
we need a larger sample to accurately measure their contribution to
the 2-8 keV CXRB.

\begin{figure}
\plottwo{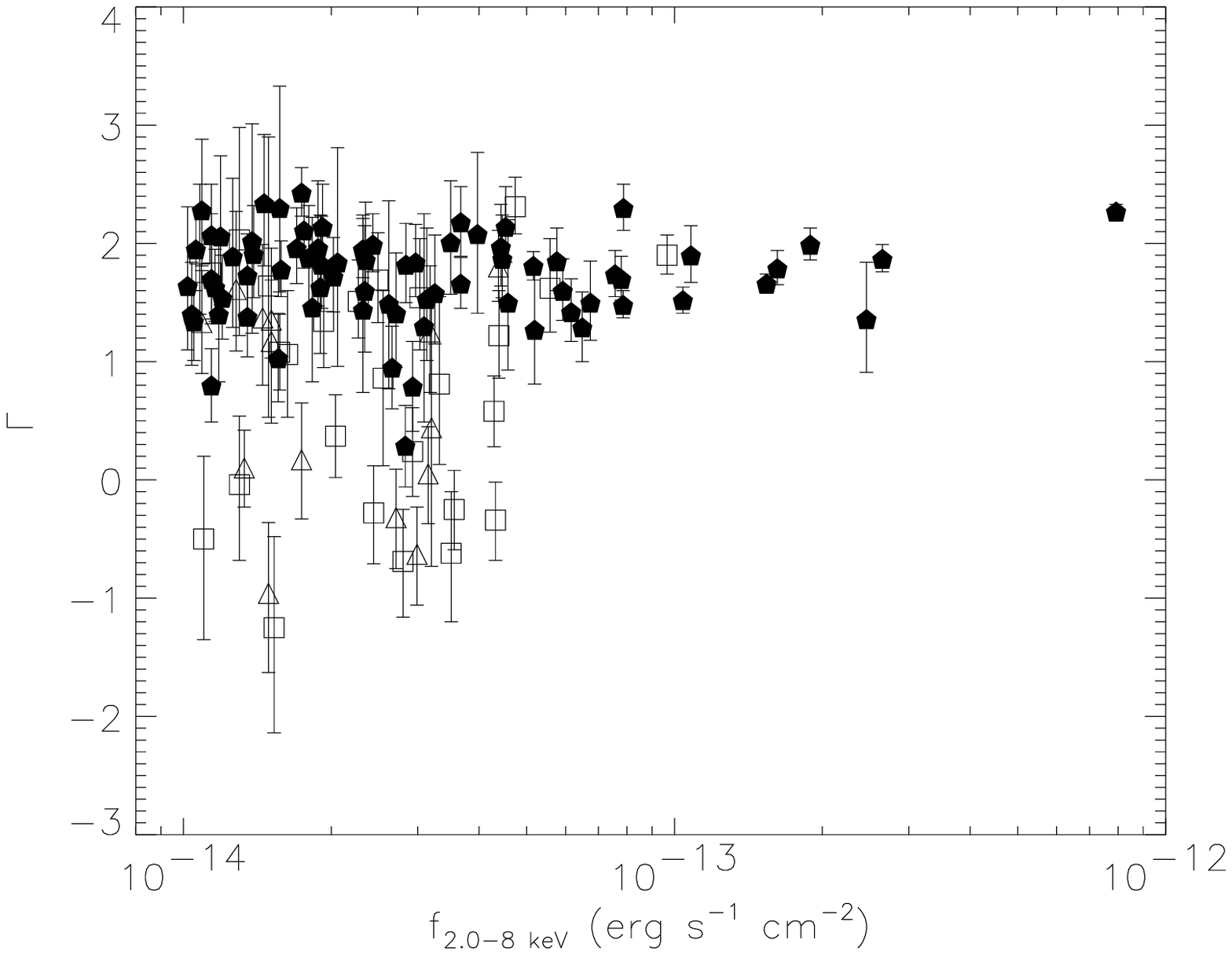}{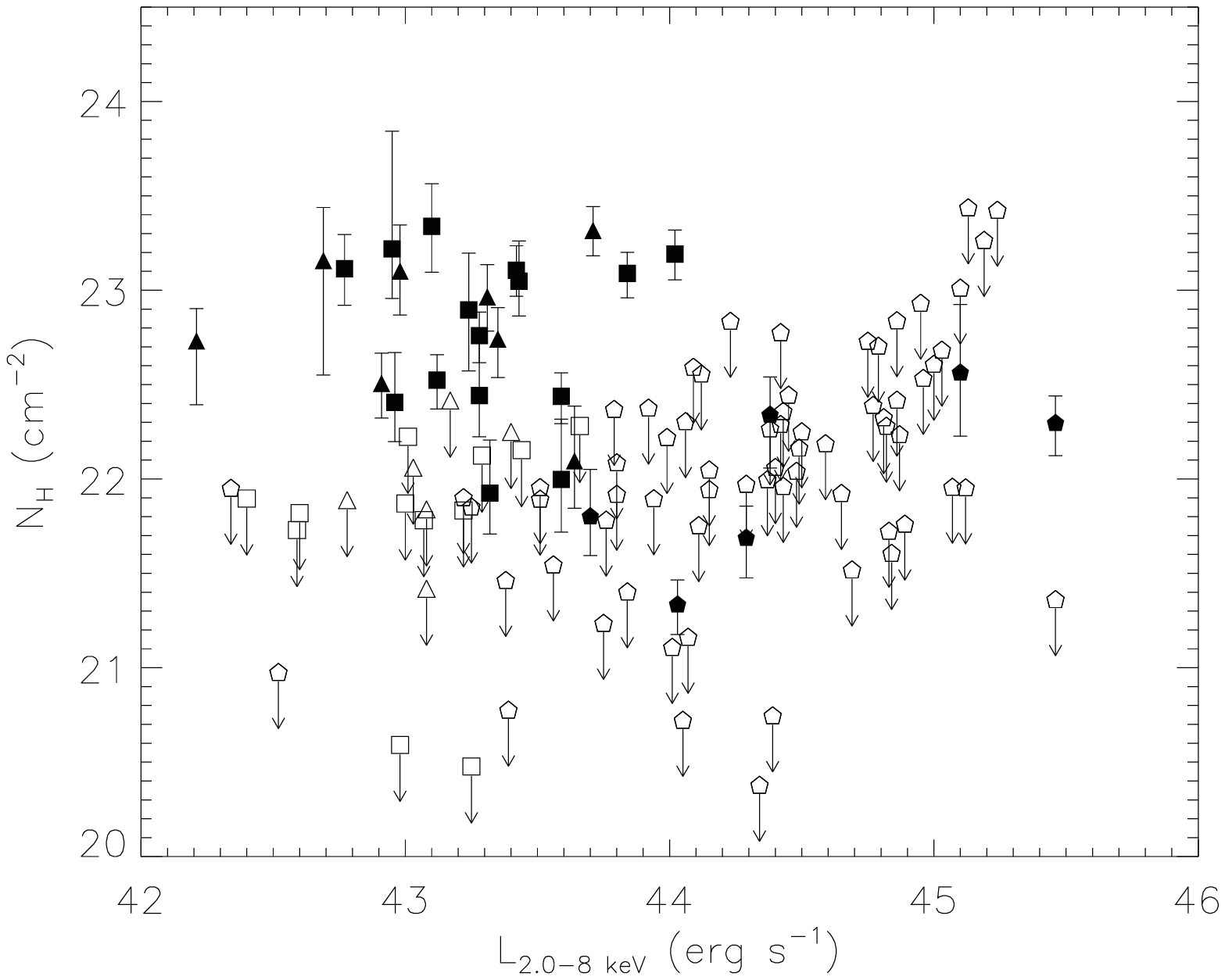}
\caption{{\it left }X-ray photon index ($\Gamma$) as a function of
X-ray flux.  {\it right} Intrinsic N$_{\rm{H}}$ vs. luminosity.
Filled symbols ({\it right}) mark the constrained measurements and
open symbols are upper limits.}
\end{figure}

\acknowledgements
We gratefully acknowledge the financial support of NASA grant:
AR2-3009X (Chandra).

\end{document}